\documentclass[acmsmall]{acmart}
\usepackage{graphics}
\usepackage{tabularx}
\graphicspath{{figures/}{pictures/}{images/}{./}} 
\usepackage{multirow}
\usepackage{microtype}
\PassOptionsToPackage{warn}{textcomp}  
\usepackage{textcomp}                  
\usepackage{mathptmx}                  
\usepackage{amsmath}
\usepackage{times}                     
\usepackage{tabu}                      
\usepackage{booktabs}                  
\usepackage{arydshln}
\usepackage{tikz}
\usetikzlibrary{calc}
\newcolumntype{b}{X}
\newcolumntype{s}{>{\hsize=.5\hsize}X}
\newcommand{\tikzmark}[1]{\tikz[overlay,remember picture] \node (#1) {};}
\newcommand{\DrawBoxA}[3][]{%
    \tikz[overlay,remember picture]{
    \draw[black,#1]
      ($(#2)+(-4.51em, 2.2ex)$) rectangle
      ($(#3)+(2.2em,-0.9ex)$);}
}
\newcommand{\DrawBoxB}[3][]{%
    \tikz[overlay,remember picture]{
    \draw[black,#1]
      ($(#2)+(-3.01em, 1.9ex)$) rectangle
      ($(#3)+(3.0em,-1.0ex)$);}
}\newcommand{\DrawBoxC}[3][]{%
    \tikz[overlay,remember picture]{
    \draw[black,#1]
      ($(#2)+(-4.51em, 2ex)$) rectangle
      ($(#3)+(1.6em,-0.7ex)$);}
}
\newcolumntype{b}{X}
\newcolumntype{s}{>{\hsize=.5\hsize}X}

\AtBeginDocument{%
  \providecommand\BibTeX{{%
    \normalfont B\kern-0.5em{\scshape i\kern-0.25em b}\kern-0.8em\TeX}}}

\setcopyright{acmcopyright}
\acmJournal{PACMCGIT}
\acmYear{2022} \acmVolume{5} \acmNumber{1} \acmArticle{5773171} \acmMonth{5} \acmPrice{15.00}\acmDOI{10.1145/3522610}

\begin{document}

\title[Effect of Resolution on Gameplay in VR]{Effect of Render Resolution on Gameplay Experience, Performance, and Simulator Sickness in Virtual Reality Games}


\author{Jialin Wang}
\affiliation{%
  \institution{Xi'an Jiaotong-Liverpool University}
  \city{Suzhou}
  \state{Jiangsu}
  \country{China}}
\email{jialin.wang17@student.xjtlu.edu.cn}

\author{Rongkai Shi}
\affiliation{%
  \institution{Xi'an Jiaotong-Liverpool University}
  \city{Suzhou}
  \state{Jiangsu}
  \country{China}}
\email{rongkai.shi19@student.xjtlu.edu.cn}

\author{Zehui Xiao}
\affiliation{%
  \institution{Xi'an Jiaotong-Liverpool University}
  \city{Suzhou}
  \state{Jiangsu}
  \country{China}}
\email{zehui.xiao16@student.xjtlu.edu.cn}
  
\author{Xueying Qin}
\affiliation{%
  \institution{Shandong University}
  \city{Qingdao}
  \state{Shandong}
  \country{China}}
\email{qxy@sdu.edu.cn}
  
\author{Hai-Ning Liang}
\authornote{Corresponding author ({\tt\small haining.liang@xjtlu.edu.cn})}
\affiliation{%
  \institution{Xi'an Jiaotong-Liverpool University}
  \city{Suzhou}
  \state{Jiangsu}
  \country{China}}
\email{haining.liang@xjtlu.edu.cn}

\renewcommand{\shortauthors}{J. Wang, R. Shi, Z. Xiao, X. Qin, and H.-N. Liang}

\begin{abstract}
 Higher resolution is one of the main directions and drivers in the development of virtual reality (VR) head-mounted displays (HMDs). However, given its associated higher cost, it is important to determine the benefits of having higher resolution on user experience. For non-VR games, higher resolution is often thought to lead to a better experience, but it is unexplored in VR games. This research aims to investigate the resolution tradeoff in gameplay experience, performance, and simulator sickness (SS) for VR games, particularly first-person shooter (FPS) games. To this end, we designed an experiment to collect gameplay experience, SS, and player performance data with a popular VR FPS game, \emph{Half-Life: Alyx}. Our results indicate that 2K resolution is an important threshold for an enhanced gameplay experience without affecting performance and increasing SS levels. Moreover, the resolution from 1K to 4K has no significant difference in player performance. Our results can inform game developers and players in determining the type of HMD they want to use to balance the tradeoff between costs and benefits and achieve a more optimal experience. 
\end{abstract}




\maketitle

\section{Introduction}
With the proliferation of mass-market head-mounted displays (HMDs), virtual reality (VR) technology has grown in popularity in the last few years. There is intensified competition among VR HMD manufacturers for hardware improvements. The HMD's resolution, often considered a measure of the level of detail provided by its two displays (one for each eye), can directly affect users' perception, performance, and experience when using the HMD. It has become one of the principal competitive factors in the rivalry among VR HMD manufacturers\footnote{A detailed comparison of all specifications is available on: \url{https://en.wikipedia.org/wiki/Comparison_of_virtual_reality_headsets}.}. Compared to other aspects such as refresh rate and field-of-view, the emphasis on resolution improvements has been moving significantly more rapidly. 

\begin{table}[tb]
  \caption{Resolution of the main consumer-level VR HMDs from 2016 to 2021. There are three distinct groups (1K, 2K, and 3K and above).}
  \label{tab:hmdTable}
	\centering
  \begin{tabu}{
	r c l
	}
  \toprule
  Device & Resolution per eye & Release date \\
  \midrule
    \tikzmark{top left 1}Oculus Rift & 1080$\times$1200 (1K) & March 2016 \\
    HTC VIVE & 1080$\times$1200 (1K) & April 2016 \\
    Sony PlayStation VR & 960$\times$1080 (1K) & October 2016 \\
    FOVE & 1280$\times$1440 (1K) & January 2017 \\
    HTC VIVE Pro & 1440$\times$1600 (1K) & January 2018 \\
    Oculus Rift S & 1280$\times$1440 (1K) & March 2019\tikzmark{bottom right 1} \\
    \tikzmark{top left 2}Oculus Quest 2 & 1832$\times$1920 (2K) & October 2020 \\
    HP Reverb G2 & 2160$\times$2160 (2K) & November 2020 \\
    Pico Neo 3 & 1832$\times$1920 (2K) & May 2021 \\
    StarVR & 2560$\times$1440 (2.5K) & April 2020 \\
    HTC VIVE Pro 2 & 2448$\times$2448 (2.5K) & June 2021\tikzmark{bottom right 2}\\
    \tikzmark{top left 3}Varjo VR-3 & 2880$\times$2720 (3K) & September 2021 \\
    PIMAX 8K & 3840$\times$2160 (4K) & February 2019 \\
    PIMAX 12K & 5620$\times$2720 (6K) & October 2021\tikzmark{bottom right 3} \\
  \bottomrule
\end{tabu}
\DrawBoxA[ultra thick, draw=white, fill=gray!100, fill opacity=0.2]{top left 1}{bottom right 1}
\DrawBoxB[ultra thick, draw=white, fill=gray!100, fill opacity=0.05]{top left 2}{bottom right 2}
\DrawBoxC[ultra thick, draw=white, fill=gray!100, fill opacity=0.2]{top left 3}{bottom right 3}
\end{table}

\autoref{tab:hmdTable} shows the resolution of the main consumer VR HMDs released in the past five years (2016 to 2021). As can be observed, 1K (approximately 1000 horizontal pixels of resolution per eye) is the dominant resolution for early consumer-level VR HMDs (e.g., Oculus Rift). There are many transitional consumer VR HMDs with a resolution between 1K and 2K (e.g., HTC VIVE Pro). 2K and above VR HMDs have not been popular until last year. Recently, many VR manufacturers have started promoting VR HMDs with a resolution above 2K (e.g., HP Reverb G2). 3K is relatively rare compared to other resolutions from 1K to 4K (e.g. Varjo VR-3). The first consumer 4K VR HMD, the PIMAX 8K (actually 4K per eye), was released in 2019 and has remained the only available 4K VR HMD in the market. The company has just announced the PIMAX 12K, the first consumer-level 6K per eye VR HMD, in October 2021 and will ship to consumers at the end of 2022. In short, improving the display resolution has always been one of the drivers in the evolution of VR HMDs.

The resolution of the human eye is 576 megapixels with a 120° field-of-view (FOV), which is roughly equivalent to 24000$\times$24000 (24K) \cite{8798011}. While 24K resolution is not within the foreseeable future, the resolution of VR HMDs will keep increasing in the next few years. However, it is unclear the benefits of VR HMDs with higher resolution for VR environments in general and for games in particular. It is also unclear if there is a correlation between resolution, simulator sickness (SS) levels, and game experience and performance. This research aims to fill this gap with a systematic user study with a popular and highly-rated VR game with four resolutions (1K, 2K, 3K, and 4K). Based on our literature review, we hypothesize that: \textit{there is a resolution threshold for VR HMDs, after which the benefits of higher resolution on SS reduction and enhanced gameplay experience become non-significant}.

The main contribution of this paper is the identification of this threshold via an empirical study. To the best of our knowledge, our experiment is the first to show this threshold for current VR HMDs. Our results show that 2K resolution is an important threshold for game experience improvement and SS reduction in VR environments, particularly in FPS games. This finding shows that, for VR HMD manufacturers, it may be more beneficial to focus on improving other aspects, like refresh rate if their device can achieve a 2K resolution, since there are no significant benefits on game experience among 2K to 4K displays. For players of VR games, they can opt for devices with 2K resolution, when the cost is an important factor or to enable other aspects of the device to run more optimally, if there are tradeoffs in performance (e.g., refresh rate, as a higher rate can reduce SS in VR applications \cite{Kourtesis2019}). Our results also show that the far-away visual details in the virtual environment in higher resolution may be the direct cause of the above difference on game experience and SS. This finding can help VR game developers to take better advantage of available high resolution to improve users' gameplay experience.

\section{Related Work} \label{RelatedWork}
In this section, we review previous research about simulator sickness in VR, resolution tradeoff, and game performance and experience.

\subsection{Simulator Sickness in VR}
Simulator sickness (SS) is still one of the biggest challenges during long time exposure of VR for a considerable number of VR users, especially for games. Previous research has shown that VR first-person shooter (FPS) games can induce more severe SS symptoms than the same FPS game played with a traditional 2D display \cite{monteiro2020depth}. An FPS game with simulated traditional 2D display in VR can even cause less SS than a normal VR stereoscopic display \cite{monteiro2020depth}. Although VR stereoscopic displays can improve depth perception and immersion, the increased SS risk still represents a serious concern for many VR users \cite{9515455,porcino2017minimizing}.

Simulator sickness questionnaire (SSQ) is the most widely used method for assessing SS \cite{Balk2013, Kennedy1993}. It can be used to quantify the level of SS for activities that may cause SS symptoms. SS is a form of motion sickness induced by the visual system's perception of movement in the environment \cite{Flanagan2004, Shi2021RacingGame, Monteiro2021Trajectory}. Three potential responses, according to the etiology, or causes, of motion sickness: reflexive eye movements, sensory conflict, and postural instability \cite{Flanagan2004}. When users interact with VR applications such as games, the eyes receive most of the stimulation. Therefore, the HMD's resolution is an important factor of VR applications. Previous research has shown that an increased resolution in a VR HMD can eliminate aspects of SS experiences in VR applications \cite{Kourtesis2019}.

\subsection{Resolution Tradeoff}
Low-resolution displays can have a negative impact on user experience. Low-resolution VR displays in an HMD may give an unclear view of the virtual environments and this can lead to unclear vision and a decreased performance and experience in VR applications. For example, in games, players may not be able to recognize items (such as types of weapons and ammo) and differentiate between enemies or team players in an FPS game. In addition, players may not be able to observe the surroundings with clarity and enjoy the views in a car racing game if it is rendered in low resolution. Finally, a low-resolution application may induce SS, which remains a difficult challenge and obstacle when long exposure and rapid movements in virtual environments are needed \cite{Rebenitsch2015, von2016cyber, Yildirim2019} and hinders the broader adoption of VR HMDs by non-regular users \cite{Diego18, McGill2017, Weech2018}. 

A high-resolution VR HMD, though it can provide more visual details of the virtual environments, requires a better hardware setup with higher performance, which might be costly for HMD manufacturers and users. Also, it can compete with other features, like refresh rate and frame rate. 
Besides the display panel, resolution, refresh rate, and frame rate are directly related to the bandwidth of the connecting cable for tethered VR HMDs and wireless transmission (wireless streaming) and performance of the dedicated GPU(s). For tethered VR HMDs, a higher resolution usually means a lower refresh rate to provide a stable video signal transmission. For wireless VR streaming, higher resolution and higher refresh rate can cause heavier load on the streaming and GPU processing, which could lead to a higher latency, and this can increase SS \cite{hecht2016optical}. Therefore, it is worth exploring how much higher resolution can benefit players during gameplay in VR environments, like games, which is an aspect that has not been explored in previous research. If a higher resolution does not help reduce SS and has no significant improvement on game experience, VR HMD manufacturers can choose display panels with lower resolution but have higher refresh rate, while VR players can choose a lower resolution setting but higher frame rate to achieve a more optimal experience.


\subsection{Game Performance and Experience}
Previous research in 2007 has shown that resolution has little impact on performance and some impact on enjoyment in non-VR games \cite{claypool2007effects}. Note that the resolution conditions (320$\times$240, 512$\times$384, 640$\times$480) from their experiment were much lower than today's consumer-level displays (typically, 1K and above). For our study, we used 4 different resolutions (1K to 4K) as the conditions for our experiment and within an immersive VR context. In this work, we also attempt to predict the benefits of higher resolution on SS reduction, which was not measured in the cited research. Moreover, they used a customized questionnaire to measure game experience since the standard questionnaire for game experience measurement, game experience questionnaire (GEQ), was published after their paper, in 2013 \cite{claypool2007effects,ijsselsteijn2013game}. 

GEQ has a modular structure that consists of 3 parts: the core questionnaire, the social presence module, the post-game module. GEQ can measure a player’s gaming experience through 7 factors: immersion, flow, competence, positive and negative affect, tension, and challenge \cite{Johnson2018}. GEQ has been widely used in different fields \cite{Boletsis2019}. A revised GEQ was proposed in a recent paper since the original GEQ has a few invalid items \cite{Johnson2018}. The revised GEQ keeps 25 items from the original GEQ and contains 5 factors: flow, immersion, competence, positive affect, negativity. To evaluate the gameplay experience in different conditions, we used the revised GEQ in our experiment.

Games play an essential, growing role in VR. However, unlike traditional 2D displays, VR games can induce more SS symptoms \cite{Ibanez2016,xu2020studying, monteiro2020depth}. FPS games are a popular genre \cite{Budhiraja2017, claypool2007effects, wu2012playing,frostling2009first}, in part because they lend themselves quite naturally for the first-person immersive view of VR. For this study, we chose a short game level from \emph{Half-Life: Alyx}, a popular and well-received VR FPS game, to collect gameplay experience, SS, and player performance data. Another reason for chosing \emph{Half-Life: Alyx} was that its photorealistic graphics may provide further insights into the effects of resolution during gameplay. The first game in the \emph{Half-Life} series has been considered a pioneer of FPS games that combine fighting enemies and solving puzzles \cite{mactavish2002technological}. \emph{Half-Life: Alyx}, the latest (2020) game in this series, also incorporates elements of both combat and puzzle-solving. Its developers place interactive objects throughout the game and players can choose to explore the virtual environment when all enemies are eliminated.

 FPS games are a popular choice as testing environment to research about games \cite{farmani2017player,krompiec2019enhanced,krekhov2017self,monteiro2021evaluating,seok2021using}. There are cases where researchers would build their own game environment (e.g., to test a new techniques or gameplay experiences \cite{Shi2021RacingGame, Xiang2021Myopic, Liu2021VRRelic, Wenge2021Uncertainty}). In our case, because we are exploring resolution, a commercial, mature, and stable gaming environment is more suitable, similar to \cite{Diego18} using \emph{Mario Kart Wii}\footnote{\url{https://www.mariowiki.com/Mario_Kart_Wii}} and \cite{Xu2021Anxiety} using \emph{FitXR}\footnote{\url{https://fitxr.com/}}. Typically, these games have good visuals and gameplay mechanisms and, in the case of \emph{Half-Life: Alyx}, come with excellent photorealistic graphics and gameplay mechanics. As such, we chose a short game level from this game as the testing environment after some technical modification to make the game fit our research purposes.

\section{User Study}
The goal of this user study was to examine the effects of render resolution on gameplay experience, performance, and simulator sickness. The scenario we selected was a VR FPS game. We conducted a within-subjects study with resolution as the independent variable. We collected participants' performance data and feedback after playing \emph{Half-Life: Alyx} with four levels of resolution, from 1K per eye to 4K per eye. This section describes the experiment conducted in this study. 

\subsection{Participants and Apparatus}
Sixteen volunteers (6 females, 10 males) were recruited from a local university. Participants' age ranged between 18 and 22 ($M=19.88, SD=1.11$). Based on the information collected from a pre-experiment questionnaire, they all had a normal or corrected-to-normal vision, had no history of color blindness, and did not declare any mental or physical health issues. Fourteen participants had some experience with 3D video games. Thirteen participants had experience with non-VR FPS games. Five participants had some experience with VR games. The experiment was conducted in accordance with the ethics guidelines and regulations of the [removed for review] and was reviewed and approved by its Research Ethics Committee. All of the volunteers consented to participate in the experiment. 

\begin{figure}[tb]
 \centering
 \includegraphics[width=0.5\columnwidth]{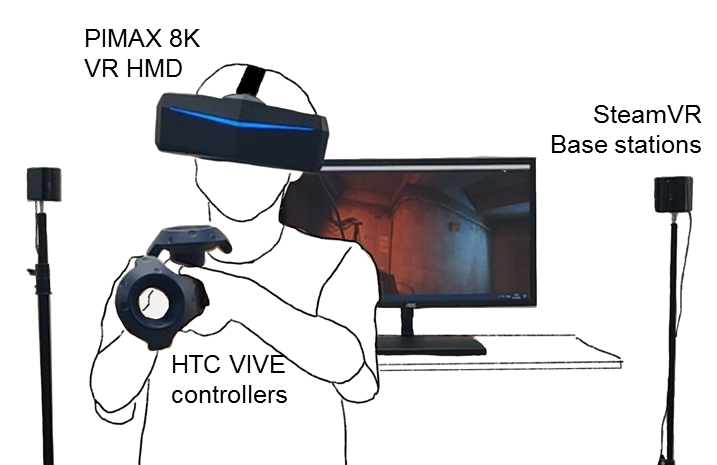}
 \caption{The experimental setup with a participant wearing the HMD and holding the controllers.}
 \label{fig:HMD}
\end{figure}

A PIMAX 8K VR HMD was used in the experiment. It is a commercial tethered VR device with an FOV of 200°(D)/ 170°(H)/ 115°(V), a refresh rate of 90Hz, and a resolution of 3840$\times$2160 per eye. The VR HMD was connected to a computer with 64 GB RAM, a GeForce RTX 3090 GPU, and an Intel Core i9-10900K CPU. Since the PIMAX 8K did not come with its own controllers, we used two HTC VIVE controllers as the input devices and two HTC SteamVR Base stations for tracking, which was the setup recommended by the company\footnote{\url{https://support.pimax.com/en/support/home}}. \autoref{fig:HMD} shows the experiment setup and the devices used in the study. 

\begin{figure*}[tb]
    \centering
    \includegraphics[width=0.8\textwidth]{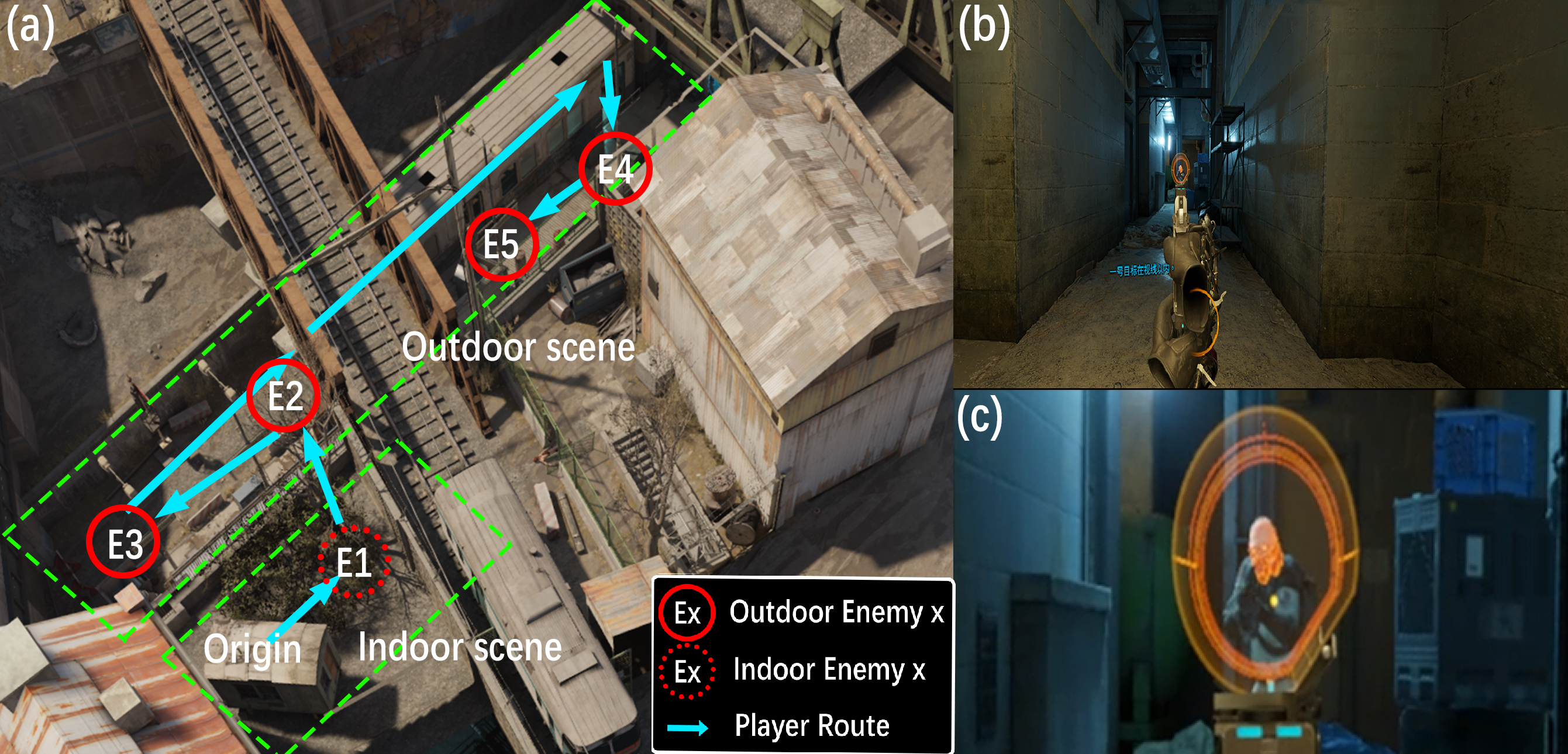}
    \caption{(a) A bird view of the game environment. (b) A screenshot of the game scene. It is from the right eye view in 4K resolution (i.e., 3840$\times$2160). The player is raising the pistol and aiming at an enemy. (c) A magnified view of the reflex sight. The enemy's weak point (head) is highlighted in orange when it is aimed at. This figure is cut and enlarged from (b); in the game, the viewpoint does not zoom in.}
    \label{fig:Game}
\end{figure*}

\subsection{Virtual Environment and Task}
We chose a game level from \emph{Half-Life: Alyx} as the testing environment for the experiment (see \autoref{fig:Game}). The actual image for each eye has a sight lens distortion effect. There were five enemy soldiers placed in different places in the game level (see \autoref{fig:Game} (a)). The game level contained an indoor scene with 1 enemy and a outdoor scene with the remaining 4 enemies. The shooting would start when an enemy spotted the player. Participants needed to defeat all 5 enemies following a route in the game (see the blue arrows in \autoref{fig:Game} (a)). \emph{Half-Life: Alyx} offers different options for locomotion or moving in or around the environment. We used the default setting of teleportation because this approach could cause less SS than linear, continuous movement \cite{monteiro2020depth} and is, in general, easier to use, especially for non-VR users. For some small-range movements, such as hiding to dodge incoming shots, participants were allowed to move physically rather than via teleport, which was a common feature of the game. We set one pistol with 999 ammo as the only weapon in the game. This was to remove the bias from different game experiences and to reduce the learning cost for the participants. Each magazine contained 10 ammo. When the ammo in the magazine ran out, participants would need to replace the magazine with a new one. The pistol had a reflex sight that provided an orange front sight (see \autoref{fig:Game} (b) and (c)). The reflex sight would support aiming at the target with the actual view instead of having an enlarged, zoomed-in view. When the participants aimed at the enemy’s weak point (head) through the reflex sight, the enemy's weak point would be highlighted in orange, as shown in \autoref{fig:Game} (c). To allow the reproducibility of the study, we provide the procedures of making the game environment in the appendix (see Section~\ref{appendix:Procedure}).

Participants were required to defeat all 5 enemies in the environment in 5 minutes. Since it would be the first time for all participants to play the game, we run a pilot study to find a suitable difficulty level. In the formal study, an enemy would be defeated if he received 2 shots to his head or 4 shots to his body. Our pilot study showed that it was possible to complete the task within the required 5 minutes after the training session. If the participants defeated all enemies in less than 5 minutes, they were told to walk around and interact with the objects in the environment. A typical participant could effectively defeat all five enemies in 3 to 4 minutes. Therefore, the gameplay consisted of a main FPS task and a short free navigation and manipulation task. This design would ensure a 5-minute gameplay in all conditions to control the exposure time, an exposure that would be long enough to lead to some level of SS from playing VR games for a typical user.


\subsection{Experimental Design}
Our study used a within-subjects design with display resolution as the independent factor. One approach to change the resolution is by replacing the display panel with different hardware displays. However, such an approach would not only lead to different weights of the HMD, which could then be a confounding factor and may also bring an increased time cost and complexity. Therefore, we decided to use render resolutions with the same display following a previous research about resolutions in an FPS game on desktop displays \cite{claypool2007effects}.

\begin{figure}[tb]
 \centering 
 \includegraphics[width=0.5\columnwidth]{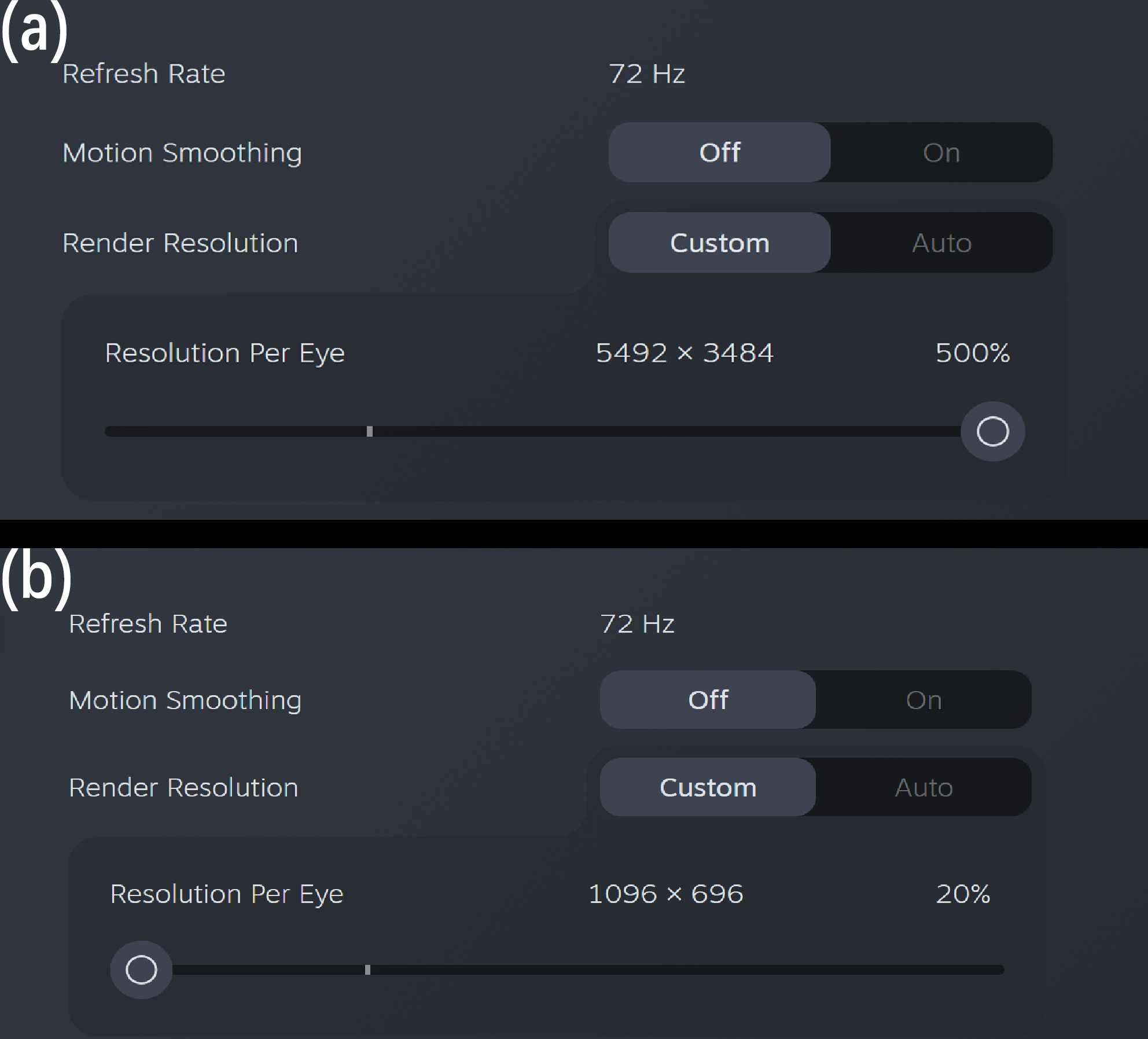}
 \caption{Screenshots of resolution per eye in SteamVR setting panel. (a) Maximum resolution per eye (super sampling: a render resolution higher than the hardware resolution of display panel); (b) Minimum resolution per eye.}
 \label{fig:Setting}
\end{figure}

\begin{figure*}[tb]
 \centering
 \includegraphics[width=0.7\textwidth]{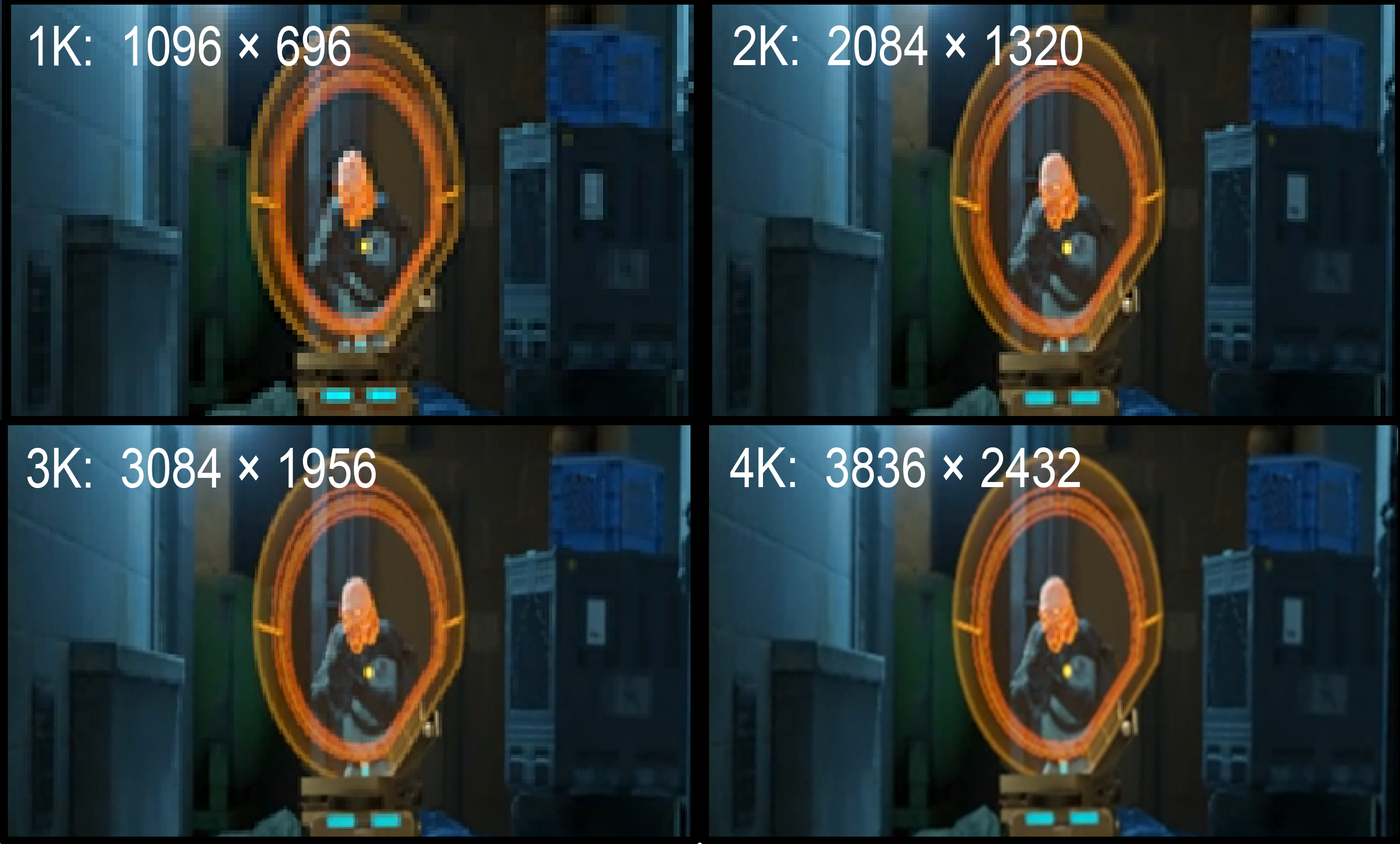}
 \caption{Screenshots of the four resolution conditions in the experiment. They are reflex sight images from right eye view and scaled by a Nearest Neighbor interpolation method for better comparison of details at the pixel level. }
 \label{fig:Conditions}
\end{figure*}

The render resolution was controlled by two rendering quality settings from PiTool (the official driver software of PIMAX 8K) and SteamVR (the driver for VR games). The final resolution was determined by the product of those two values. We setup four resolutions for the experiment: 1096$\times$696 (1K), 2084$\times$1320 (2K), 3084$\times$1956 (3K), and 3836$\times$2432 (4K). A render resolution higher than the hardware resolution of display panel is so-called super sampling (see \autoref{fig:Setting}). Although it is good for improving the game experience, we still used 3836$\times$2432 as the highest resolution condition since super sampling is out of the aim of this research. It could be a variable to explore its effects on game experience in future studies. The highest resolution was close to the hardware limit (3840$\times$2160). As shown in \autoref{tab:hmdTable}, this represented the highest resolution available in today's commercial HMDs and, as such, the four conditions (1K to 4K) would cover them. \autoref{fig:Conditions} shows the reflex sight images of 4 the conditions used in the experiment. The enemies showed at the same positions among all conditions since the game level used in the experiment is a linear level (without random elements).

Although the hardware of the displays could support 4K with 90Hz, PIMAX 8K has a bottleneck on the video signal transmission with a high refresh rate in 4K resolution\footnote{\url{https://community.openmr.ai/t/pimax-the-8k-will-have-90-hz-or-75-hz-refresh-rate-with-a-1070/3830/4}}. We contacted the company and one of its engineers explained that the two displays would produce unstable renderings with shifting in high refresh rates (above 72Hz) in 4K resolution due to issues related to HDMI and DP interface. Because of this technical constraint, we had to use a refresh rate of 72Hz in the experiment. Moreover, the frame rate of the game was limited to 90 by the RTSS of MSI Afterburner. Finally, the graphics quality was set to Ultra to ensure photorealistic graphics, which may reveal more effects of resolution during gameplay.

Although the FOV is adjustable, we used full FOV (200°(D)/ 170°(H)/ 115°(V)) to minimize any negative effect of a small FOV on the gameplay experience \cite{rakkolainen2016superwide}. We also tested the HMD device with the game to ensure that there were no other technical issues. For example, the wide FOV of PIMAX 8K could cause some rendering issues in many VR games due to camera culling. Rendering issues were fixed before the experiment using a solution found on the official development community website of PIMAX 8K \footnote{\url{https://community.openmr.ai/t/how-to-get-rid-of-rendering-issues-in-half-life-alyx-without-parallel-projection/29004}}. 

Except for the resolution, all other aspects were kept constant in all 4 conditions. In other words, we controlled all other variables (e.g., testing environment, game difficulty, graphics quality, refresh rate, frame rate, and FOV) to explore only the effects of resolution.

\subsection{Measurement}
We used the simulator sickness questionnaire (SSQ) to measure the levels of sickness and the revised game experience questionnaire (GEQ) to measure the gameplay experience \cite{Kennedy1993, Johnson2018}. SS was measured by the post-exposure SSQ questionnaire after each condition. The results were further processed to derive four sub-scores: Total Severity, Nausea, Oculomotor, and Disorientation. We considered the relative SSQ scores ($\Delta$SSQ) which were calculated by subtracting the SSQ scores before a trial from the SSQ scores after this trial. The revised GEQ had five sub-scores: flow, immersion, competence, positive affect, and negativity \cite{Johnson2018}. 

In addition to the subjective measurements, we also collected participants' gameplay performance in each condition. For each condition, we counted the number of fired ammo, and the number of hits received. 

\begin{figure*}[tb]
 \centering
 \includegraphics[width=\textwidth]{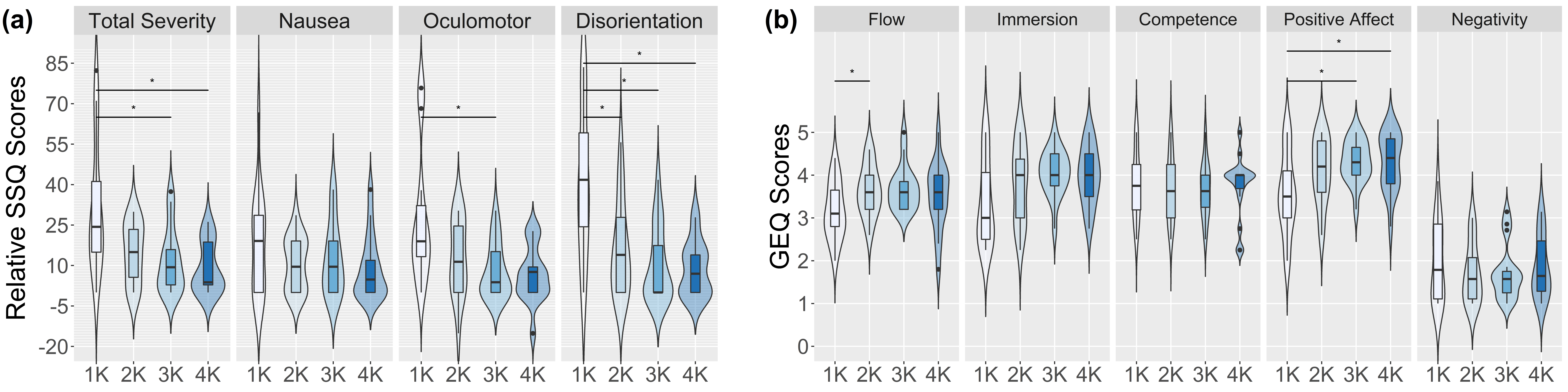}
 \caption{Violin and box plots of (a) four relative simulator sickness questionnaire (SSQ) sub-scores, and (b) five game experience questionnaire (GEQ) sub-scores. `*' represent a `0.05' significance level with Bonferroni correction. }
 \label{fig:SSQGEQ}
\end{figure*}

\begin{figure}[tb]
 \centering
 \includegraphics[width=0.4\columnwidth]{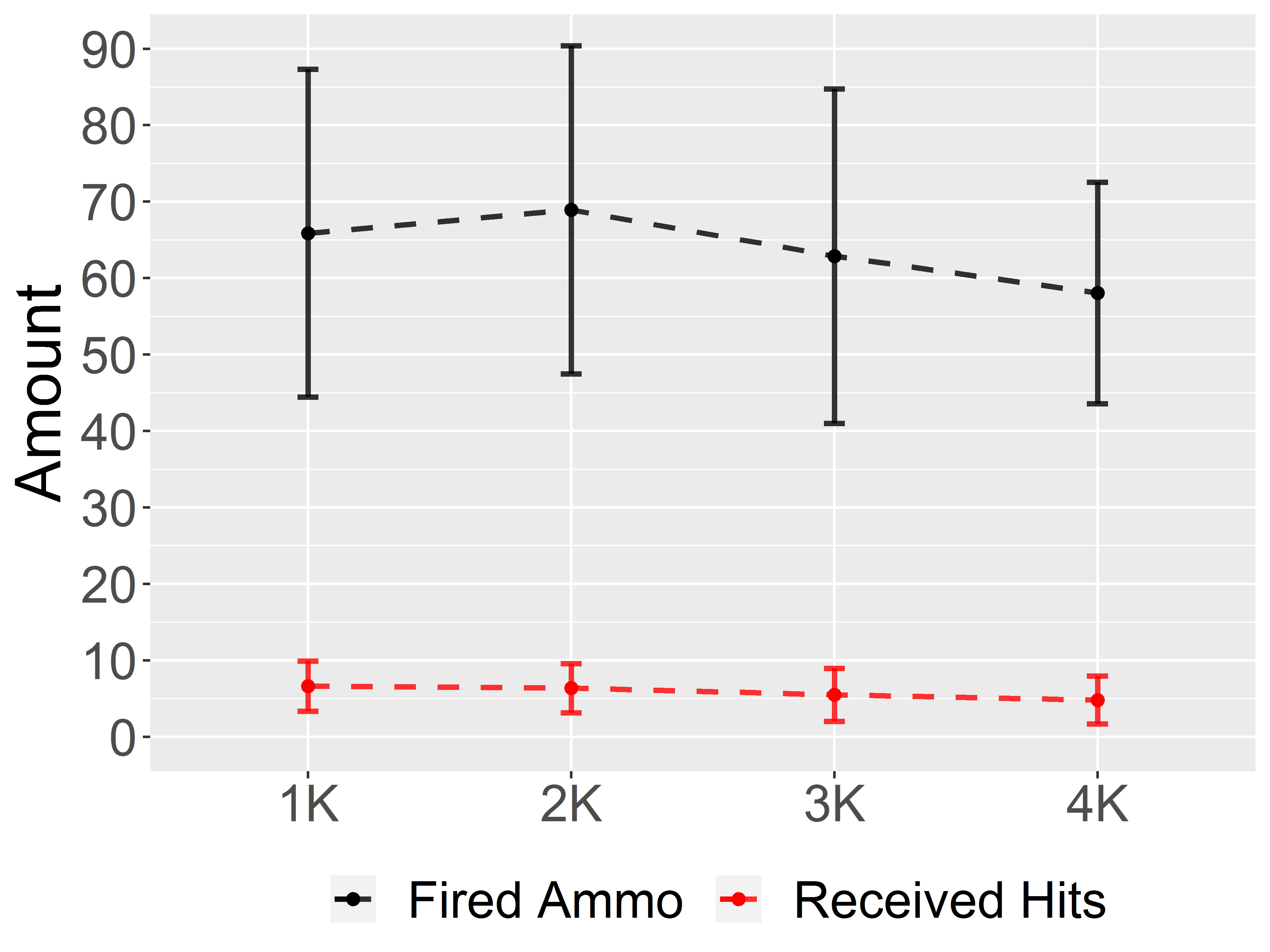}
 \caption{Mean amount of ammo fired and hits received. Error bars represent 95\% confidence intervals.}
 \label{fig:Player}
\end{figure}

\subsection{Procedure}
Each participant was assigned a specific order of conditions in which he or she would play. The order of the conditions was determined through a balanced 4$\times$4 Latin-Square design to mitigate carry-over effects. Participants were first invited to fill in a questionnaire to collect demographic information and past VR and gaming experience before the experiment. After, they were required to watch a 5-minute tutorial video of the game. After adjusting the IPD of the VR HMD, participants were given some time to learn to use VR controllers to control movement, aim, shot, and reload in a 5-minute training session. 
Before the formal experiment started, participants had enough rest to prevent the accumulation of SS and filled in a pre-SSQ. Then the participants began the formal experiment. They needed to wear external earphones to ensure an integrated gaming experience since the PIMAX 8K has no internal earphones. After each condition, the participants were required to fill in a post-SSQ and a revised GEQ. The gameplay of each condition was recorded by the game for analysis of each participant's performance. The game save was replaced after each condition to reset the testing environment. We gave the participants enough rest between two conditions. The participants could stop at any point if they felt uncomfortable during the experiment. There was a short interview about their perception of the different resolutions and general experience after the participants finished all conditions. The whole study lasted about 50-60 minutes.

\section{Results}
We used SPSS version 25 for Windows for data analysis. We performed non-parametric Friedman tests for data collected via the SSQ and GEQ. Post-hoc analysis with Wilcoxon signed-rank tests was conducted with Bonferroni corrections, resulting in significance levels at 0.00833 (0.05/6), 0.00166 (0.01/6), and 0.00016 (0.001/6). On the other hand, we used a parametric repeated measures ANOVA (RM-ANOVA) for player performance data. For each measurement, we first conducted analyses to investigate the effects of order of the conditions.

\subsection{Simulator Sickness}
Results from a Friedman test showed that the order of the conditions did not significantly affect SSQ sub-scores ($p>0.05$). \autoref{fig:SSQGEQ} (a) shows the results of relative SSQ sub-scores. The Friedman test revealed statistically significant differences in Total Severity ($\chi^2\left(2\right)=17.234, p=0.001$), Oculomotor ($\chi^2\left(2\right)=11.446, p=0.010$), and Disorientation ($\chi^2\left(2\right)=20.500, p=0.001$) among the four resolution conditions. Post-hoc pairwise comparisons showed significant differences when comparing the Total Severity scores of 1K, 3K, and 4K (3K-1K: $Z=-2.952, p=0.003$, 4K-1K: $Z=-2.873, p=0.004$) in favour of 3K and 4K (lower Total Severity scores), as well as the Oculomotor scores of 1K and 3K (3K-1K: $Z=-2.640, p=0.008$) in favour of 3K (lower Oculomotor scores), and the Disorientation scores of 1K, 2K, 3K, and 4K (2K-1K: $Z=-2.848, p=0.004$, 3K-1K: $Z=-3.089, p=0.002$, 4K-1K: $Z=-2.996, p=0.003$) in favour of 2K, 3K and 4K (lower Disorientation scores). We did not find significant difference in the Nausea scores ($\chi^2\left(2\right)=6.086, p=0.108$) among the different resolutions.

\subsection{Game Experience}
Results from the Friedman test showed that the order of the conditions had a significant effect on competence scores ($\chi^2\left(2\right)=12.354, p=0.006$) but not on the remaining GEQ sub-scores ($p>0.05$). Post-hoc analysis revealed that the competence scores in the third condition was significantly higher than in the first condition ($Z=-2.884$, $p=0.004$). \autoref{fig:SSQGEQ} (b) shows the results of the revised GEQ sub-scores. The results of the Friedman test indicated statistically significant differences in the flow scores ($\chi^2\left(2\right)=8.764, p=0.033$) and positive affect scores ($\chi^2\left(2\right)=15.205, p=0.002$) among the different resolutions. Post-hoc analysis revealed significant differences when comparing the flow scores of 1K and 2K (2K-1K: $Z=-2.711, p=0.007$) in favour of 2K (higher flow scores), as well as the positive affect scores of 1K, 3K, and 4K (3K-1K: $Z=-3.020, p=0.003$, 4K-1K: $Z=-2.949, p=0.003$) in favour of 3K and 4K (higher positive affect scores).

The results of the Friedman test did not indicate any statistically significant differences in the immersion scores ($\chi^2\left(2\right)=4.015, p=0.260$), competence scores ($\chi^2\left(2\right)=0.583, p=0.900$), and negativity scores ($\chi^2\left(2\right)=1.007, p=0.799$) among the different resolutions.

\subsection{Player Performance}
RM-ANOVA revealed that the order of the conditions had an significant effect on the number of hits received ($F\left(3, 1\right)=3.271, p=0.030$) but not on the amount of ammo fired ($F\left(3, 1\right)=2.212, p=0.100$). In contrast, post-hoc analysis did not show any significant differences ($p>0.00833$). \autoref{fig:Player} summarizes the results of player performance. An RM-ANOVA with Greenhouse-Geisser correction ($\epsilon<0.75$) did not indicate any statistically significant differences in the amount of ammo fired ($F\left(2.651, 39.764\right)=0.878, p=0.450$) and the number of hits received ($F\left(2.128, 31.914\right)=0.461, p=0.647$) among all resolution conditions.

\subsection{General Feedback}
After finishing all the conditions of the experiment, we conducted a semi-structured interview asking participants about their game experience or other feelings about the game in the four different resolutions, especially with 1K. Ten participants explicitly mentioned that it was hard to have a clear vision of the target in the game in 1K. Six of them highlighted that they had difficulty recognizing the figures of the enemies, and found it hard to aim at them even with the support of the reflex sight and having the color hints. Two participants (P11, P12) said, `\textit{Due to the low resolution, the orange hint on the reflex sight blurred the view and made it harder to aim at the target.}' Ten participants reported some degree of difficulty to observe the surroundings in the game clearly in 1K. 
Two participants (P10, P15) stated that they needed more response time in lower resolutions.

\section{Discussion}
The results of SSQ analysis indicate that 2K, 3K, and 4K can produce significantly lower SS symptoms than 1K at a refresh rate of 72Hz. The results of GEQ analysis indicate that 2K can help the participants to enter into a significantly deeper flow state than 1K. Moreover, 3K and 4K can produce a significantly stronger positive affect than 1K. However, there are no significant differences among 2K, 3K, and 4K for each factor of both SSQ and GEQ. The player performance analysis shows that there are no significant differences in the number of ammo fired and hits received among 1K, 2K, 3K, and 4K. We discuss these findings further below. 

Regarding SSQ, Oculomotor is related to the level of disturbance of visual processing during simulation (eye disturbance level) \cite{Kennedy1993}. The Oculomotor scores given by participants show that their eye disturbance level in 3K could have been significantly lower than in 1K. The p value for 4K-1K is also close to 0.00833 (0.011 to be exact). One inference from this result is that higher resolution could be beneficial in reducing the level of participants' eye disturbance because of the possibility to see details with more clarity in high resolution. Similarly, Disorientation scores given by participants to 2K, 3K, and 4K are significantly higher than to 1K, which may imply that the participants could get disoriented more easily in a low-resolution setting (like 1K, due to the difficulty to see more details).  

\begin{figure}[tb]
 \centering 
 \includegraphics[width=0.7\columnwidth]{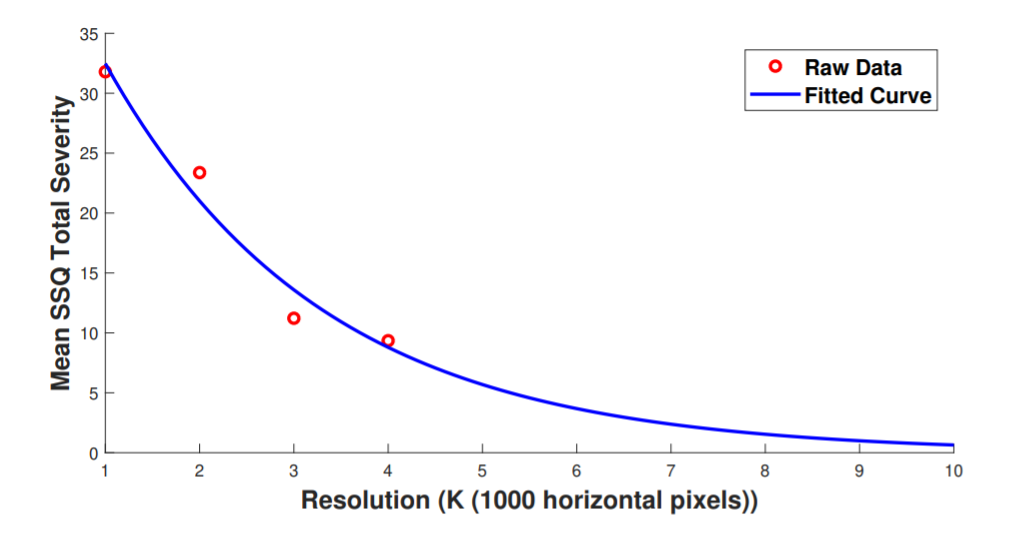}
 \caption{An Exponential Model of the mean Total Severity scores (from SSQ) vs. resolutions.}
 \label{fig:Model}
\end{figure}

Total Severity from SSQ is a more general metric for VR games and other similar applications. We found that the mean Total Severity of the four conditions could be fitted into an exponential model ($f\left(x\right) = 50.21*e^{-0.4356*x}$) with the sum of squared error (\textit{SSE}) of 12 and a coefficient of determination ($R^2$) of 0.9643. The fitted curve shows that the benefits of resolution on SS reduction would exponentially decreases as it becomes higher (see \autoref{fig:Model}). As can be observed, the steepest part is between 1K to 2K, after which the curve flattens rapidly. This also shows that the displays near the 2K resolution represent a significant threshold, after which there are diminishing returns. This finding implies that developers could focus on a 2K display and strive to have higher frame or refresh rates, as higher resolution (than 2K) with the same frame rate as in our experiment would not bring that many benefits, especially in SS reduction and enhanced gameplay. There could be differences in SS and game experience among 2K, 3K, and 4K and above for other types of games. On the other hand, given that we used an FPS game that requires a fast action-reaction time, the observation from the model could apply to other similar types of VR games and applications. 


The interview about participants' general experience in lower resolution may provide hints to the benefits of high resolution. Participants commented that seeing details clearly in games is the most important concern when it comes to resolution. The front view is one of the most critical areas of focus for games in general but specifically for FPS games. Although most participants mentioned that the front view was sometimes hard to see clearly in lower resolution (especially 1K), all resolution conditions did not lead to significant differences in participants' performance. However, for VR games that rely more on observing minute details (e.g., pure puzzle games, and to some extent, role-playing games), the higher the resolution the better the experience could likely be. These games would typically be slower paced, which would also mean that players would likely have more time to make detailed observations of the objects in the virtual environment. This additional time could also be a compensatory factor for lower/higher resolution.

The above results all lead to the relationship between resolution and level of detail. Moreover, 2K, 3K, and 4K can provide enough details to have a significantly better result than 1K in most metrics. There is no significant difference among 2K, 3K, and 4K in all metrics. The 4 conditions also have no significant difference in player performance. These results provide strong support for our hypothesis mentioned in the introduction (that there is a limit of benefits from VR HMDs with higher resolution on SS reduction and game experience improvement, at least in VR FPS games). The above results can lead an important conclusion: 2K is the threshold value for resolution that can ensure a good experience with fewer SS symptoms and without affecting player performance for VR FPS games.

This is a helpful and important finding that can perhaps serve as a guide for VR HMD manufacturers, players, and game developers to enhance users' gameplay experience. We have mentioned earlier that there is a tradeoff between resolution and refresh rate and between resolution and frame rate in the related work section (see Section~\ref{RelatedWork}). Higher resolution, refresh rate, and frame rate can potentially cause lower SS symptoms \cite{matsushima201851,brennesholtz20183}. Therefore, VR HMD manufacturers can choose 2K displays with a higher refresh rate since 2K is the threshold value for the resolution that can ensure a suitable overall experience. A 2K display with a higher refresh rate can further reduce SS symptoms. This is similar to frame rate, which can be adjusted by VR players. VR games are GPU-heavy applications that require a tradeoff between high resolution and high frame rate for both tethered VR HMD and wireless VR streaming. Hence, VR players can choose 2K resolution with higher frame rate. In addition, developing games with 2K resolution tends to be practical for developers since 2K resolution does not require art assets with very high quality. In general, for both VR HMD manufacturers and VR players, 1K should be avoided to ensure an enjoyable level of game experience with fewer SS symptoms and a desirable level of performance, at least in VR FPS games.

\section{Limitations and Future Work}
This section discusses the limitations of this research, which could represent directions for future work. First, we only tested the different resolutions in an FPS game. The testing environment and type of game are two limitations of the current work. We chose \emph{Half-Life: Alyx} after considering several other games, including in-house games developed by our group. At the end, \emph{Half-Life: Alyx} was selected because the game was representative of current VR games in the market and suited well the topic of this research in that its game experience might be highly related to the resolution. \emph{Half-Life: Alyx} is a popular game with photorealistic graphics and, as our results show, using it in this research provided further insights into the effects of resolution during gameplay. The results can be a reference of VR games that need the benefit of high resolution (e.g., those with photorealistic graphics). However, not all games may require the perception of details brought by high resolution such as \emph{Beat Saber}. Comments by participants during the interview suggest that VR games that require careful observation of visual details may still need higher resolution. This hypothesis needs to be confirmed in future studies with other types of games (like role-playing games) to explore if a higher threshold resolution is beneficial for such games, which would typically be slower in pace. Second, although our current work suggests that 2K is a significant threshold, the accurate threshold value may be somewhere between 1K and 2K. As such, it is worth exploring a more exact threshold (or range) since many VR HMDs still use displays with resolutions between 1K and 2K. In addition, our experiment fixed the refresh rate, frame rate and field-of-view to allow us to focus on resolution. On the other hand, the experiment can be reproduced to explore the benefits of (lower or higher) refresh rate and frame rate and (wider or narrower) field-of-view on SS reduction, gameplay experience, and performance in VR FPS and other types of games and applications. Finally, while our experiment has produced significant results, it has involved a relatively homogeneous group (in terms of VR/game experience, age, and cultural background). In the future, it will be helpful to run similar experiments involving a larger and more diverse group to see if the same results hold and what other insights are observed that can guide the design and development of VR games to provide players more optimal experiences without incurring in higher costs.  

\section{Conclusion}
In this paper, we present an experiment and analysis for resolution tradeoff in game experience and simulator sickness (SS) reduction in virtual reality (VR) games. We used a modified game level from \emph{Half-Life: Alyx} as the testing environment to explore one hypothesis derived from our literature review: there is a resolution threshold for VR HMDs, after which the benefits of higher resolution on SS reduction and enhanced gameplay experience become non-significant. The analysis of experiment results showed that 2K, 3K, and 4K have a significantly better game experience and significantly fewer SS symptoms than 1K at a refresh rate of 72Hz. There are no significant differences in SS and game experience among 2K, 3K, and 4K resolutions. There are also no significant differences in the player performance among all conditions (from 1K to 4K). Our results suggest that 1K should be not used whenever possible, and that 2K is the threshold resolution that can ensure a good experience with lower SS symptoms but without any adverse effect on player performance for VR games.

\appendix
\begin{acks}
The authors wish to thank the participants who joined the study and the reviewers for their insightful comments and helpful suggestions that helped to improve our paper. This work was supported in part by Xi'an Jiaotong-Liverpool
University--Key Special Fund (\#KSF-A-03) and the Future Network Scientific Research Fund (\#FNSRFP-2021-YB-41).
\end{acks}

\bibliographystyle{ACM-Reference-Format}
\bibliography{sample-sigconf.bbl}
\appendix 
\label{appendix:Procedure}
\section{Procedure of reproducing the game environment}
The save files were made using the following steps. First, we need to enable in-game console commands before starting a new game. Second, the following commands need to be entered in the in-game console: \verb|sv_cheats 1|, \verb|map a2_train_yard|; and after to walk to the next weapon shop (the origin of testing environment, see \autoref{fig:Game}). Third, we need to type the following commands in the in-game console: \verb|give item_hlvr_weapon_energygun|, \verb|hlvr_energygun_grant_upgrade 1|, \linebreak \verb|hlvr_addresources 999 0 0 0|. Finally, we need to change the difficulty to story mode before finding and saving a copy of the save files. 

\end{document}